\documentclass[twocolumn,showpacs,preprintnumbers,amsmath,amssymb]{revtex4}

\usepackage{graphicx}
\usepackage{dcolumn}
\usepackage{bm}
\usepackage{epsf}

\usepackage{amsmath}
\usepackage{pstricks}

\DeclareSymbolFont{AMSb}{U}{msb}{m}{n}
\DeclareMathSymbol{\N}{\mathbin}{AMSb}{"4E}
\DeclareMathSymbol{\Z}{\mathbin}{AMSb}{"5A}
\DeclareMathSymbol{\R}{\mathbin}{AMSb}{"52}
\DeclareMathSymbol{\Q}{\mathbin}{AMSb}{"51}
\DeclareMathSymbol{\I}{\mathbin}{AMSb}{"49}
\DeclareMathSymbol{\C}{\mathbin}{AMSb}{"43}

\newcommand{\be}{\begin{equation}}
\newcommand{\ee}{\end{equation}}
\newcommand{\bea}{\begin{eqnarray}}
\newcommand{\eea}{\end{eqnarray}}

\newcommand{\parti}[1]{\frac{\partial}{\partial #1}}
\newcommand{\ehoch}[1]{\exp\!{\left[ #1 \right]}}
\newcommand{\ew}[1]{\langle #1 \rangle}
\newcommand{\partdiff}[2]{\frac{\partial #1}{\partial #2}}
\newcommand{\tr}[1]{\operatorname{tr}\!\left\{ #1 \right\}}
\newcommand{\kreuz}{^{\dagger}}
\newcommand{\T}{\text{\scriptsize T}}

\begin{document}

\title{System--bath entanglement in nanothermodynamics}

\author{Stefanie Hilt$^{1,2}$}
\author{Eric Lutz$^{2}$}
\affiliation{$^{1}$Institute of Quantum Physics, Ulm University, 89069 Ulm, Germany\\
$^{2}$Department of Physics, University of Augsburg, 86135 Augsburg, Germany}

\date{\today}

\begin{abstract}
We consider a quantum harmonic oscillator linearly coupled to a bath of harmonic oscillators and evaluate  the degree of entanglement between system and bath using the negativity as an exact entanglement measure. We establish the existence of a critical temperature above which the system--bath negativity vanishes. Our results imply that system--bath entanglement is not responsible for the violation of the Clausius inequality observed in the low temperature/strong coupling regime  [Phys. Rev. Lett. {\bf 85}, 1799 (2000)], as the latter still occurs well above the critical temperature.
\end{abstract}
\maketitle

Thermodynamics is a cornerstone of modern physics. In the canonical description, it deals with  systems that are in contact with a heat bath at constant temperature. An implicit assumption in the notion of thermal contact is that the system--bath interaction is vanishingly small, so that the total energy is simply the energy of the system plus that of the bath \cite{kub68}. This approximation is well--justified in the limit of high temperatures, when the thermal energy of the system is much larger than the coupling energy. In the low--temperature limit, however, quantum--mechanical effects come into play and the  interaction between the system and the bath cannot be neglected anymore \cite{han05}. It has recently been shown \cite{all00,all02} that close to zero--temperature, a finite coupling strength leads to a violation of the familiar Clausius inequality, $\delta {\mathcal Q}\leq T d S$, which states that the infinitesimal heat  $\delta {\mathcal Q}$ exchanged with the bath cannot exceed the product  of the temperature $T$ of the bath and the infinitesimal entropy change $d S$. Accordingly, work could be extracted from a single quantum heat bath. A scheme to experimentally verify  this violation using nanoscale electrical circuits has been put forward in Ref.~\cite{all02a}. So far, the  origin of the above deviation from standard thermodynamics in the quantum regime has  been associated with entanglement between  system and  heat bath \cite{all00,all02,all02a,hoe05,kim06,kim07,lef03}. However, this conjecture has never been tested explicitly.  

In this paper, we compute the system--bath entanglement using a microscopic model consisting of a quantum harmonic oscillator linearly coupled to a chain of harmonic oscillators (Rubin model) \cite{rub60}. Due to the linearity of the model, its dynamics is exactly solvable. Moreover, since its thermal states are Gaussian, we can rely on exact entanglement measures like the negativity \cite{vid02}. An early study of  entanglement between system and bath using the PPT criterion for  $1\times N$ Gaussian modes \cite{wer01} has been performed in Ref.~\cite{eis02}. 
In the following, we numerically investigate the dependence of the  system--bath negativity on the coupling strength and the temperature, and compare it to the deviation from the Clausius inequality. We find that the negativity vanishes  above a critical temperature and that the Clausius inequality is violated even when system and bath are separable. Our results clearly demonstrate that the violation is not solely induced by system--reservoir entanglement.  

{\it Thermodynamics in the quantum regime.} The starting point of our analysis is the total Hamiltonian,
\begin{align}
\label{eq1}
	H = H_S + H_B + H_I \ ,
\end{align}
which describes a system $S$ coupled to bath $B$ via the interaction $I$. 
In classical thermodynamics, the interaction energy is assumed to be small and  therefore neglected. However, for very low temperatures, the thermal energy of the system is not necessarily much larger than the interaction energy and the coupling term cannot simply be discarded. This is particularly relevant for solid--state nanodevices, such as nanomechanical oscillators, which can be strongly coupled to their environment \cite{rou01}. For finite coupling strength, the system Hamiltonian  does not in general  commute with the total Hamiltonian, $[H_S,H]\neq 0$.  Since there are then  no joint eigenstates of $H_S$ and $H$, a system strongly coupled to a zero--temperature  bath is unavoidably in an excited state, in stark contrast to normal thermodynamics \cite{lin84}. Consequently, the system can perform positive work by relaxing to its ground state. It is important at this point to emphasize that the second law is not violated, since the produced work cannot exceed the work initially  required to couple system and bath \cite{for06,kim06,kim07}.

In order to perform a quantitative study of the violation of the Clausius inequality, we consider the Rubin model   of a heavy harmonic oscillator coupled to a closed chain of harmonic oscillators \cite{rub60}. This model has played a major role in the microscopic understanding of Brownian motion \cite{zwa01}. The total Hamiltonian of the Rubin model is of the form \eqref{eq1} with:
\begin{align}
\label{eq2}
	H_S= \frac{p^2}{2  M}+\frac{ M \omega_S^2}{2}\, q^2\ ,
\end{align}
\begin{align}
\label{eq3}
	 H_B=\sum_{\alpha=1}^N\left[\frac{p_\alpha^2}{2 m}+\frac{ m \omega_B^2}{2}\,x_\alpha^2\right]
	+\sum_{\alpha=1}^{N-1}\frac{f}{2}\big(x_{\alpha+1}-x_\alpha\big)^2 \ ,
\end{align}
\begin{align}
\label{eq4}
	H_I= \frac{f}{2}\Big[(q-x_1)^2+(x_N-q)^2\Big]\ .
\end{align}
For simplicity, the $N$ oscillators of the chain have identical mass $m$ and frequency $\omega_B$. The respective mass and frequency of the system are denoted by $M$ and $\omega_S$, and the coupling strength is given by the spring constant $f=m \omega_R^2/4$. The Rubin model is fully characterized by the spectral density function ($\omega<\omega_R$),
\be
\label{eq5}
J(\omega) = \frac{m}{2}\sqrt{\omega^2+\omega_R\omega_B} \, \sqrt{(\omega_R-\omega_B)\omega_R-\omega^2} \ .
\ee
In the limit of large system mass, $M\gg m$, and small bath frequency, $\omega_B\ll \omega_R$, the Rubin model reduces to a Caldeira--Leggett-type model with Ohmic dissipation, $J(\omega)= \gamma M\omega \Gamma_D^2/(\omega^2+\Gamma_D^2)$, with friction coefficient $\gamma = m \omega_R/2 M$ and Debye cutoff frequency \mbox{$\Gamma_D=  \omega_R$ \cite{zwa01}}.
System and  bath are furthermore supposed to be initially decoupled with a total density operator given by  $\rho(0)=\rho_S(0) \otimes \rho_B$ \mbox{and $\rho_B = \exp(-\beta H_B)/Z_B$ with $\beta=(kT)^{-1}$.}

The dynamics of the system is conveniently described using the phase--space representation. The Wigner function  of the system is found to satisfy the   equation \cite{all00,hu92},
\begin{align}
\label{eq6}
	\parti{t}&W(q,p,t)=-\frac{p}{M}\parti{q}W+\parti{p}\Big(\left[\gamma p+M \omega_\text{S}^2q\right]W\Big)\nonumber\\
	&+\gamma M D_{pp}(q,t)\,\frac{\partial^2}{\partial p^2}W+\frac{\partial^2}{\partial q\partial p}\Big(D_{qp}(q,t)\,W\Big) \ .
\end{align}
In the high--temperature limit, the generalized diffusion coefficient $D_{qp}$ vanishes and Eq.~\eqref{eq6} reduces to the usual Klein--Kramers equation \cite{zwa01}. 
The stationary solution of Eq.~\eqref{eq6} is given by a quasi--Gibbs distribution,
\begin{align}
\label{eq7}
	W(q,p)=\frac{1}{2\pi}\sqrt{\frac{\omega_S^2}{T_pT_x}}\ehoch{-\frac{p^2}{2MT_p}-\frac{M \omega_{S}^2q^2}{2T_q}} \ ,
\end{align}
with effective temperatures defined as $T_q= M\omega_\text{S}^2 \ew{{q}^2}$ and $T_p=\ew{{p}^2}/M$. The  stationary position and momentum  quadratures can be calculated explicitly using Eq.~\eqref{eq5} in the Ohmic regime and read \cite{gra84},
\begin{align}
	\ew{{q}^2}
	&=\frac{T}{M\omega_\text{S}^2}+\frac{\hbar}{M\pi}\sum_{i=1}^3\left[\frac{(\lambda_i-\Gamma_\text{D})\;\psi\!\left(1+\frac{\beta\hbar\lambda_i}{2\pi}\right)}{(\lambda_{i+1}-\lambda_i)(\lambda_{i-1}-\lambda_i)}\right] \ ,\\
	\ew{{p}^2}
	&=M^2\omega_\text{S}^2\ew{\hat{q}^2}+\nonumber\\&\frac{M\hbar\gamma\Gamma_\text{D}}{\pi}\sum_{i=1}^3\left[\frac{\lambda_i\;\psi\!\left(1+\frac{\beta\hbar\lambda_i}{2\pi}\right)}{(\lambda_{i+1}-\lambda_i)(\lambda_{i-1}-\lambda_i)}\right]\ .
\end{align}
In the above equations,  $\lambda_i$ are the characteristic frequencies of the damped harmonic oscillator  and $\psi$ denotes the digamma function. In the limit of high temperatures/weak coupling, the two temperatures become equal, $T_q=T_p=T$, and Eq.~\eqref{eq7} reduces to the usual Gibbs distribution. However, in the opposite limit of low temperatures/strong coupling, the system is  squeezed by the coupling to the bath and $T_q< T_p$. As a result, deviations from standard thermodynamics appear.
 
\begin{figure}
\centering
\includegraphics[width=0.4\textwidth]{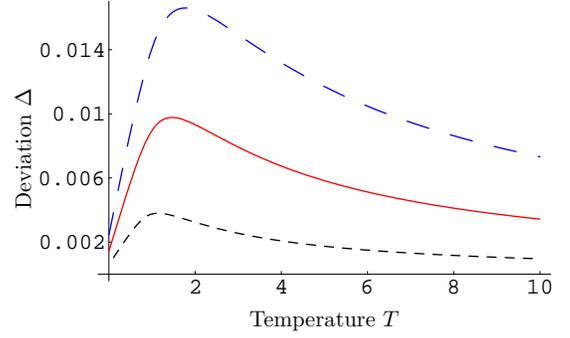}
\caption{(Color online) Deviation from the Clausius equality, $\Delta =\delta {\mathcal Q}-T  dS$, for increasing values of  the friction  coefficient $\gamma$:  $\gamma=0.3$ (black narrow dashed),  $\gamma=0.6$ (red solid) and $\gamma=0.9$ (blue dashed), for the  parameters $M=10$, $m=1$, $\omega_\text{S}=5$, $\hbar=1$ and $k=1$.}\label{fig1}
\end{figure}

We next focus on the violation of the Clausius inequality in the low temperature/strong coupling regime. The internal energy $U$ of the system is defined as the stationary expectation value of its energy, that is,
\begin{align}
	U&=\ew{{H}_S}=\frac{\ew{{p}^2}}{2 M}+\frac{M\omega_{S}^2}{2}\ew{{q}^2} \ .
\end{align}
A quasistatic  variation of a system parameter then yields
\begin{align}
	 dU&=\iint dqdp\; W(q,p)\, dH_S(q,p)+\nonumber\\
	&\iint dqd p\;H_S(q,p) \, dW(q,p)\ .
\end{align}
The first term on the right hand side is identified with the  work $\delta\mathcal W$  done on the system and the second term with the heat $\delta \mathcal Q$ exchanged with the bath.
In the following, we consider a variation of the mass $M$ of the system. The change in heat is accordingly \cite{all00,all02},
\begin{align}
	{\delta {\mathcal Q}}&=\left(\frac{M\omega_S^2}{2}\partdiff{\ew{{q}^2}}{M}+\frac{1}{2 M}\partdiff{\ew{{p}^2}}{M}\right) dM\,.
\end{align}
On the other hand, the von Neumann entropy $S$ of the system is given by \cite{all00,all02},
\begin{align}
	S&=-\tr{\rho_S\ln\rho_S}\nonumber\\
	&=\big(v+\frac{1}{2}\big)\ln \big(v+\frac{1}{2}\big)-\big(v-\frac{1}{2}\big)\ln\big(v-\frac{1}{2}\big) \ ,
\end{align}
where $\rho_S$ is the stationary density operator of the system corresponding to  Eq.~\eqref{eq7} and $v=\sqrt{\ew{{q}^2}\ew{{p}^2}}/\hbar$. It is useful to introduce the quantity $\Delta = {\delta {\mathcal Q}}-T{ d S}$, where $dS$ is defined as $d S=\partial S / \partial M dM$.  The Clausius inequality then implies that  $\Delta\leq 0$, the equality being only achieved for quasistatic transformations \cite{kub68}.
Figure \ref{fig1} shows the deviation $\Delta$ as a function of the temperature of the bath for different values of the friction coefficient $\gamma$. We clearly recognize that the Clausius inequality is violated, $\Delta>0$, for low temperatures, the violation being stronger the larger the value of the coupling $\gamma$. On the other hand, with increasing temperature the deviation tends to zero, as expected for the quasistatic mass variation that we examine. It is commonly believed that this violation of the Clausius inequality in the quantum domain is due to the entanglement created between the system and the bath as a result of their interaction \cite{all00,all02,all02a,hoe05,kim06,kim07,lef03}. We now show that this assumption is not correct.

{\it System--bath entanglement.} Quantifying bipartite entanglement for general mixed states is a non--trivial task. However, for the special class of Gaussian states, such as the thermal oscillator states of the Rubin model, exact entanglement measures do  exist. In the present investigation, we use the negativity  to characterize system--bath entanglement. 
The negativity is defined as \cite{vid02},
\begin{align}
\label{eq14}
	{\mathcal N}(\rho)= \frac{|\!|\rho^{T_S}|\!|-1}{2} \ ,
\end{align}
where $\rho$ is the total density operator of system plus bath and $\rho^{T_S}$ denotes the partial transposed with respect to the system. The trace norm of an operator $A$ is given by $|\!|A|\!| = \tr{\sqrt{AA\kreuz}}$. The negativity is equal to the sum of the modulus of the negative eigenvalues of $\rho^{T_S}$ and therefore quantifies the degree of non--positivity of this operator. The negativity is zero for separable states and increases with increasing degree of entanglement. 

The Gaussian states of the Rubin model are fully characterized by the covariance matrix, 
\begin{align}
\label{eq15}
	\Gamma_{jk}(\sigma)=\frac{1}{2}\ew{\xi_j \xi_k+\xi_k\xi_j}_{\sigma}-\ew{\xi_j}_{\sigma}\ew{\xi_k}_{\sigma} \ ,
\end{align}
where ${\xi}=(q,p,x_1,p_1,\ldots,x_N,p_N)^\T$ is the vector of  the combined positions and momenta of the system and bath oscillators. The symbol $\ew{\mbox{ } }_{\sigma}$ represents the expectation value taken with respect to a density operator $\sigma$.
The trace norm $|\!|\rho^{T_S}|\!|$ in Eq.~\eqref{eq14} can then be directly expressed in terms of  the symplectic eigenvalues $\nu_j$ of the covariance matrix $\Gamma(\rho^{T_S})$ \cite{vid02}:
\begin{align}
	|\!|\rho^{T_S} |\!|=\prod_{j=1}^{N+1}\left\{\begin{matrix}
		1	&\text{for}& \nu_j\geq 1/2\,\\
		1/2\nu_j&\text{for}&  \;\nu_j<1/2
             \end{matrix}\right.\,,
\end{align}
The symplectic eigenvalues of the matrix $\Gamma(\rho^{T_S})$ are the eigenvalues of the product  $\Omega\Gamma(\rho^{T_S})$, where $\Omega$ is the symplectic matrix; they occur in conjugate pairs $i\nu_j,-i\nu_j$. 
For the case of $(1\times N)$ Gaussian modes that we consider, the negativity can  be written in the simple  form \cite{ser06},
\begin{align}
\label{eq17}
	{\mathcal N}(\rho)=\left\{\begin{array}{ccc}
	\left(\frac{1}{4\nu_\text{min}}-\frac{1}{2}\right)&\text{for}&\nu_\text{min}<1/2\\
		0			&\text{for}&\nu_\text{min}\geq1/2
           \end{array}\right. \ ,
\end{align}
with $\nu_\text{min}=\operatorname{min}\{\nu_j\}$ the smallest symplectic eigenvalue.

\begin{figure}
\centering
\includegraphics[width=0.4\textwidth]{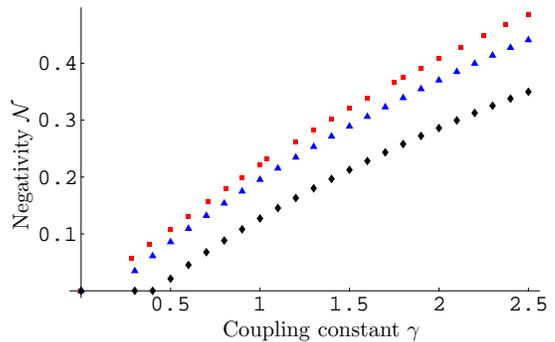}
\caption{(Color online) System--bath negativity $\mathcal N$ as a function  of the coupling constant $\gamma$ for different temperatures: $T=0.1$ (red boxes), $T=1$ (blue triangles) and $T=2$ (black diamonds), for the parameters $ M=10, m=1, \omega_\text{S}=5$, $\omega_\text{B}=0.01$,  $\hbar=1$ and  $k=1$.}\label{fig2}
\end{figure}

\begin{figure}
\centering
\includegraphics[width=0.4\textwidth]{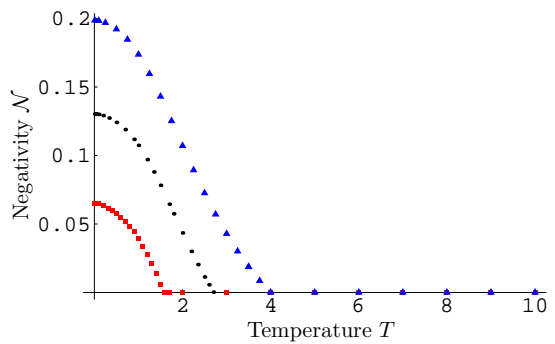}
\caption{(Color online) System--bath negativity $\mathcal N$ as a function  of  the bath temperature $T$ for different coupling constants: $\gamma=0.3$ (red boxes) $\gamma=0.6$ (black dots) and $\gamma=0.9$ (blue triangles), for the same  parameters as above.}\label{fig3}
\end{figure}

{\it Numerical results.} We have numerically computed the system--bath negativity,  Eq.~\eqref{eq17}, and examined its dependence on the coupling constant $\gamma$ and bath temperature $T$. To this end, we have evaluated the full time evolution of the coordinate vector $\xi(t)$ and determined the covariance matrix \eqref{eq15} using the  Williamson normal form of the Hamiltonian \cite{wil36,sim99}; the system was taken to be initially in the ground state. The symplectic eigenvalues $\nu_j$ of the partially transposed covariance matrix  were then computed in the stationary regime. The results of our numerical calculation for a bath of $N=200$ oscillators are summarized in Figs. \ref{fig2} and \ref{fig3}.

Figure \ref{fig2} shows the negativity as a function of the coupling constant $\gamma$  for various bath temperatures. As expected, entanglement between system and bath increases with increasing coupling strength. However, for high temperatures, the existence of a critical coupling intensity $\gamma_c$ below which the negativity is zero is clearly visible. The presence of a threshold is further confirmed in Fig.~\ref{fig3} showing the  temperature dependence of the negativity  for different coupling constants. We observe that  system--bath entanglement decreases with increasing temperature and completely vanishes above a critical temperature $T_c$. The value of the critical temperature grows with growing  coupling strength. 

It is instructive to compare our findings with the recent entanglement phase diagram proposed by Anders and coworker for a closed chain of identical oscillators in thermal equilibrium \cite{and08,and08a}; the Hamiltonian  of  the latter is given by Eqs.~\eqref{eq2}--\eqref{eq4} with $M=m$ and $\omega_S=\omega_B$.   Entanglement between adjacent oscillators  in the identical chain was shown to vanish at   a temperature $kT_c= \hbar\omega_R/2\sqrt2$, or in our case, $kT_c = \hbar\gamma M/2m$. This estimate of the critical temperature, or equivalently of the critical coupling, $\hbar\gamma_c= 2m kT/M$, for a fixed temperature, agrees very well with the values found for  the Rubin model in Figs.~\ref{fig2} and \ref{fig3}: $kT_c\simeq 5 \hbar\gamma$ for $M=10$ and $m=1$. We therefore obtain the important result that system--bath entanglement disappears at approximately the same temperature as entanglement {\it within} the bath.
 
\begin{figure}
\centering
\includegraphics[width=0.4\textwidth]{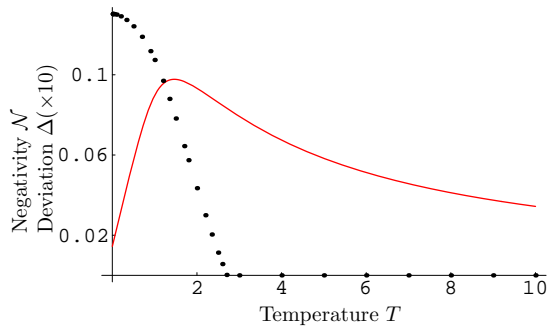}
\caption{(Color online) Deviation from the Clausius equality $\Delta$ (red line) and system--bath negativity $\mathcal N$ (black dots) as a function of the bath  temperature $T$  for $\gamma=0.6$. All other  parameters are those of Figs~\ref{fig1}-\ref{fig3}. The violation of the Clausius equality is significant above the critical temperature at which the negativity abruptly drops to zero.}\label{fig4}
\end{figure}

{\it Violation of the Clausius inequality.} The relationship between the violation of the Clausius inequality and the presence of system--bath entanglement can now be established by combining the results of Figs.~\ref{fig1} and \ref{fig3}. Figure \ref{fig4} unambiguously shows that the deviation, $\Delta>0$, from the Clausius inequality persists at temperatures way above the critical temperature at which the negativity vanishes. We can  therefore conclude that the deviation from ordinary thermodynamics is not caused by   the entanglement of the system with the bath, contrary to what has been claimed up to now. We note in addition that this discovery has also deep consequences for the theory of decoherence, where it is often believed that the loss of phase coherence originates from the entanglement of the  system with the external reservoir \cite{giu96}. Again, this  cannot be the case for temperatures above $T_c$. We stress that the negativity can already vanish in the regime of low temperatures, $kT_c<\hbar\omega_S$, as illustrated in Fig.~\ref{fig3}. 

To summarize, we have performed a detailed study of the entanglement between system and bath in the Rubin model of a heavy oscillator coupled to a chain of harmonic oscillators and showed that it vanishes above a critical temperature $T_c$. We have further provided compelling evidence that the violation of the Clausius inequality which occurs in the regime of low temperature/strong coupling is not induced by system--bath entanglement, as it persists well above $T_c$.

This work was supported by the Emmy Noether Program of the DFG (Contract LU1382/1-1) and the
cluster of excellence Nanosystems Initiative Munich (NIM). We thank W. Schleich and M. Freyberger for their support and A.  Allahverdyan for discussions.

\end{document}